\documentclass[11pt,twoside]{article}
\usepackage{txfonts}\usepackage{balance}\usepackage{arxiv}\usepackage{natbib}
\usepackage{graphicx}
\usepackage{float}
\usepackage[a4paper]{hyperref}

\newcommand{\hinode}{\emph{Hinode\ }}
\begin{document}

\title{Solar chromospheric flares: energy release, transport and radiation}
\author{Lyndsay~Fletcher
\affil{School of Physics and Astronomy, SUPA, University of Glasgow, Glasgow G12 8QQ, UK.}
}

\begin{abstract}
This paper presents an overview of some recent observational and theoretical results on solar flares, with an emphasis on flare impulsive-phase chromospheric properties, including: electron diagnostics, optical and UV emission, and discoveries made by the \hinode mission, especially in the EUV. A brief perspective on future observations and theoretical requirements is also given.
\end{abstract}

\section{Introduction}
During a solar flare, magnetic energy that was stored in stressed coronal magnetic field is released to dramatic effect. The principal result is the intense burst of radiation  that occurs across the electromagnetic spectrum,  in association with (and possibly caused by) copious numbers of charged particles accelerated out of a thermal background. Also frequently occurring at the time of a solar flare is a coronal mass ejection: the magnetically-driven expulsion of large quantities of plasma, and its entrained magnetic field, into the heliosphere. 

Explaining flare particle acceleration requires explaining how the energy stored on macroscopic scales (in twisted or stressed or tangled field) is imparted to individual particles, which - if they do not travel into the heliosphere -  go on to produce heating, excitation and radiation in the lower solar atmosphere. Clearly the radiation produced in the flaring atmosphere is the primary means of understanding and diagnosing this process. Flare radiation is dominated by the optical and UV part of the spectrum \citep{1989SoPh..121..261N,2004GeoRL..3110802W,2006JGRA..11110S14W}, which - from imaging observations - are of chromospheric origin, indicating that the chromosphere is the primary energy-loss region and the primary source of diagnostic radiation. So it is here that we must focus our attention. The optical and UV part of the spectrum has been somewhat neglected in recent flare studies. However the few observations that there are, those made with the \hinode Solar Optical Telescope (SOT) coupled with those in EUV and X-rays, have proved to be very informative about the condition and response of the chromospheric plasma. 

This article focuses on diagnostics of the flare plasma, but of course a flare cannot be understood without looking at the state and the evolution of the magnetic field. With its capacity for measuring the photospheric vector field, the \hinode Solar Optical Telescope has proved very informative in this respect too.

\section{Models}\label{sect:models}

The majority of flare energy release takes place during the first few minutes of a flare, in the impulsive phase. Later on, particularly in so-called long duration events (LDEs) when the field has simplified into a clear arcade or cusp structure, there is evidence from the anomalously long-lived hot coronal plasma that substantial energy release may continue throughout the gradual phase \citep{1995ApJ...446..860F,2006ApJ...638.1140J,2011A&A...531A..57K}. Nevertheless it is the impulsive phase that proves the most serious challenge both in terms of the complexity of the magnetic field involved and the rate of energy release.
In standard models of the impulsive phase, the conversion of stored magnetic energy is primarily into the kinetic energy of non-thermal particles, and takes place in the corona, usually in close association with magnetic reconnection.  The acceleration of charged particles on the Sun requires an electric field, which may be an ordered, large-scale field such as in a current sheet or shrinking magnetic structure, or a stochastic field such as in a turbulent outflow region (see \cite{2011SSRv..159..357Z} for a recent theoretical overview). The emphasis of acceleration models is to identify the origin of that field and to describe how particles are accelerated in the field, evaluating the typical energies and (more rarely) also the  number fluxes obtainable \citep{2006SoPh..236...59H,2011A&A...528A.104M}, this being the more difficult property to explain.  Flare energy should, according to observations, be transported quickly to the chromosphere to explain the tendency to  simultaneity between the non-thermal hard X-ray (HXR) footpoint emission \citep{1996AdSpR..17...67S,2002SoPh..210..229K}, and since it is clear from non-thermal HXR emission - generated by the electron-proton bremsstrahlung mechanism - that there are copious numbers of electrons present in the chromosphere, which is also strongly heated, a beam of high energy particles from the corona kills three birds with one stone: transporting the energy, producing the HXRs, and collisionally heating the lower atmosphere. 

An alternative view is that, rather than being primarily converted in the corona to the KE of non-thermal particles, energy propagates away from the reconnection site as a magnetic disturbance - i.e. Poynting flux \citep{2008ApJ...675.1645F,2009ApJ...695.1151B}, which dissipates wherever plasma conditions favour it. The location and nature of the dissipation will depend on the nature of the disturbance and on the magnetic and plasma properties of the environment through which it propagates. This view promotes the idea that the magnetic `convulsion' of the CME propagating upwards is accompanied by a magnetic convulsion propagating downwards, with roughly the same total energy (as indicated by studies of flare and CME energetics) but launched into a much smaller volume and onto a field which is line-tied at the photosphere. This Poynting-flux picture is far less developed than the electron beam model. 

The most direct diagnostic for fast electrons during flares is HXR emission (and the $\gamma$-ray emission for ions, though this is more difficult to observe). The majority of HXR emission comes from the chromosphere, though both thermal and non-thermal coronal sources are also frequently observed. The standard model hypothesis is that the coronal and chromospheric HXR emission is generated by parts of a single electron distribution, with the coronal emission produced by accelerated electrons as they pass through, or are trapped in, the coronal field, and the chromospheric emission by `precipitating' electrons from the same distribution \citep{1976MNRAS.176...15M}. Alternatively the two populations could be distinct but both produced, albeit in different locations, by the flare magnetic disturbance. In the framework of the standard flare model, the implied electron beam flux from the corona can be as high as a few $\times 10^{36}{\rm electrons~s^{-1}}$ \citep[e.g.][]{1976SoPh...48..197H,2003ApJ...595L..97H}. The small spatial scale of the footpoints (HXR as well as their optical counterparts)  implies a beam number density of up to $10^{10}{\rm cm^{-3}}$ \citep{2011ApJ...739...96K}, which is a challenge. In one case of a flare in which the footpoints were behind the limb, \cite{2010ApJ...714.1108K} reported evidence that all electrons in the coronal source of density $\sim 2 \times 10^9 {\rm cm^{-3}}$ were accelerated (note, coronal field convergence would increase the beam density at the chromosphere, but would also reduce the precipitating fraction, unless the beaming along the field were very strong).
However, usually, the coronal HXR fluxes, and thus the number of accelerated coronal electrons, are lower by a factor of around 5-10 \citep{2008ApJ...673.1181K}, so in this sense the chromospheric and coronal HXR source requirements calculated from the normal collisional propagation models do not `match up'. It should also be noted that even in normal coronal sources, the inferred non-thermal electrons can be numerically significant (energetically dominant) and the usual notion of a thermal `core' with a superposed non-thermal `tail' distribution having a distinct low-energy cut-off may not be helpful, despite being a convenient distribution to use for spectral fitting.

\section{Electron diagnostics}

Flare HXR images and spectra, dominated by the chromospheric emission, give quite direct diagnostics of the flare electrons at high energies. The property most readily obtained from HXR footpoint measurements, once the bremsstrahlung emission cross-section has been deconvolved, is an averaged source property - the density-weighted, source-averaged electron spectrum. By then assuming a particular model for the behaviour of the electrons as they propagate, further inferences can be made. In particular, by assuming the `collisional thick target' model, which says that electrons enter the top of the chromosphere and are collisionally degraded as they propagate until they merge with the thermal background, the parameters of the beam injected at the top of the chromosphere can be deduced in a straightforward manner. If processes other than collisional stopping happen - for example deceleration in self-induced electric fields \citep{2006ApJ...651..553Z} or wave-particle interactions which may lead to modified energy losses or indeed energy gains  \citep{2006GMS...165..157M,2011arXiv1112.4448K}, then the injected beam properties are not obtained so straightforwardly. Modifications to the HXR spectrum by photopheric Compton backscattering must also be corrected for \citep{1978ApJ...219..705B,2006A&A...446.1157K}

Chromospheric HXR observations tell us that at energies above about 25~keV the (density-weighted, source-averaged) electron spectrum is roughly a power-law. A power-law is characterised by its spectral index $\delta$, its intensity, and (to keep the energy content finite) its low-energy cutoff $E_c$. From HXR observations, the first two can be determined, but only an upper limit can be set on the third. Without a better determination of the low-energy cutoff the total flare energetics remains poorly constrained. For example, as the total energy in a power-law spectrum depends on $E_c^{(2-\delta)}$ the difference between a low energy cutoff of 20~keV and 10~keV leads to a factor 8 difference in total energy, for $\delta = 5$. So getting an idea of the low-energy cutoff is very important for total flare energetics, and it may be that other chromospheric diagnostics can play a role here. However, given the complexity of the chromospheric response, involving ionisation, heating, hydrodynamics, and the generation of both optically thick and optically thin radiation, the diagnostics will have to be accompanied by detailed modeling.

How might the electron parameters be constrained by other observations? The heating of the solar atmosphere, and its effect on the optically thin and thick radiation thus produced, offers several routes. In the framework of the cold collisional thick target model, there is a simple relationship between the column depth $N = \int n dl$ through which an electron of a given energy penetrates before losing all of its energy, and the initial energy of that electron \citep[e.g.][]{1978ApJ...224..241E}. Low energy electrons stop high in the atmosphere, and high energy electrons stop deep in the atmosphere. The stopping location is also where the electron gives up the bulk of its energy. An electron distribution with a low value of cutoff energy will (especially if it has a relatively steep spectrum) primarily heat the upper chromosphere, and vice versa. This would be reflected in the character of the chromospheric evaporation, and the footpoint differential emission measure.
Flare heating, ionization and collisional excitation also affect the spectral line profiles of lines formed throughout the atmosphere, to different extents depending on location and distribution of temperature and density and hence energy input. For example the temperature-height structure around the temperature minimum region can be studied using the Ca II K lines \citep{1978SoPh...58..363M}, and the temperature and density structure respectively from  Mg I at 4571\AA\ and 5173\AA\  \citep{1990ApJ...350..463M,1990ApJ...365..391M}. The mid and upper chromosphere are sampled by other lines such as the Mg I b multiplet \citep{1993ApJ...414..928M}, CIV 1549\AA\ and of course the hydrogen Balmer series. Both the structure of the heated flare atmosphere \citep[e.g.][]{1984ApJ...282..296C} and the parameters of the electron distributions, whose direct collisional contributions affect the hydrogen line formation \citep{1993A&A...274..917F,1993SoPh..143..259Z,2002A&A...382..688K}, are reflected in the Balmer line profiles. 

Finally, flare radio emission, particularly that corresponding to gyrosynchrotron radiation from non-thermal electrons in strong magnetic fields, as is relevant in the chromosphere, offers rich diagnostic potential for flare electrons. For a review of this extensive field see \cite{2011SSRv..159..225W}

\section{Chromospheric flare sources}
Morphologically, chromospheric flares appear as extended bright ribbons visible in the UV, EUV, H$\alpha$ and occasionally other parts of the optical spectrum. A subset of locations on these ribbons are also host to HXR footpoint sources \citep[e.g.][]{2002ApJ...578L..91A}. When observations have allowed - i.e. when stable, high-cadence data are available - HXR footpoints are seen also to have optical (`white light') counterparts \citep{2007ApJ...656.1187F}. The optical and HXR sources are very well correlated in space and time. HXR sources are also spatially co-incident with the strongest H$\alpha$ sources \citep{2002ApJ...578L..91A}. Not so well-known is the fact that HXR footpoint sources are accompanied by impulsive soft X-ray footpoint sources \citep{2004A&A...415..377M} which suggest temperatures on the order of 8-10~MK, and densities of a few times $10^{10}{\rm cm}^{-3}$, located rather low in the atmosphere (i.e. not coronal loops, but also not deep in the chromosphere, judging by the densities.) 

Flare impulsive phase optical spectroscopy is these days mostly restricted to the H$\alpha$ lines, though occasionally the Ca II line at 8542\AA\ \citep{2006ApJ...653..733C} and other metal lines are observed \citep[e.g. some basic spectroscopy has been carried out with the Fe I line at 6173\AA\ used by SDO/HMI by ][]{2011SoPh..269..269M}. In previous decades, other metal lines were studied giving insight into e.g. the behaviour of the temperature minimum region during flares, but such studies are no longer commonly done. UV spectroscopy of chromospheric flares on the disk has not been carried out since the end of the 1980s,  but in the 70s and 80s spectroscopic observations were made with Skylab \citep{1977ApJ...215..329D} and OSO-8 \citep{1984SoPh...90...63L} and spatially-resolved observations of flare kernels were obtained with the scanning slit UV Spectrometer and Polarimeter (UVSP) on the Solar Maximum Mission enabling, for example, density diagnostics \citep{1981ApJ...248L..39C}. Serendipitous off-limb observations of the flare UV spectrum scattered by the corona have also been made using two instruments, the Solar Ultraviolet Measurement of Emitted Radiation (SUMER) instrument \citep{2004A&A...418..737L} and the UltraViolet Coronagraph Spectrometer (UVCS) instrument \citep{2007ApJ...659..750R}. Of course, both the Transition Region and Coronal Explorer (TRACE) and the Solar Dynamics Observatory (SDO) make high-resolution imaging measurements in certain UV bands (1600\AA, 1700 \AA) which encompass important chromospheric lines.  However, the UV behaviour remains relatively unknown and it is to be hoped that the forthcoming IRIS mission will greatly improve on the current state of knowledge.

On the other hand, the EUV part of the spectrum is now being well exploited to learn about the temperature, density and velocity conditions in the hotter part of the lower atmosphere in flares (see Section~\ref{sect:EUV}). It is a little difficult to know what to term this region of the flaring atmosphere; the densities are chromospheric and the morphology of the sources consistent with footpoints and ribbons, but the temperatures at which these diagnostic lines are formed are typical of the corona, at around 1 to 1.5~MK, and possibly higher.
 
\subsection{Hard X-ray and optical footpoints}

A recent volume reviewing the results from the RHESSI spacecraft and their associated theoretical developments gives a complete overview of what has been accomplished over the last 9 years in the field of flare hard X-ray observations \citep{2011SSRv..159....1E}. Consistent with what was seen by the {\it{Yohkoh}} Hard X-ray telescope, RHESSI typically revealed small numbers of compact non-thermal flare footpoints, appearing roughly simultaneously (though RHESSI's imaging relies on having at least a half-spin of the spacecraft, taking 2 seconds, meaning that simultaneity is difficult to examine closely) and co-spatially with the brightest parts of the flare ribbons. Occasionally also extended ribbons of HXR emission are seen but in general the footpoint distribution indicates an irregularly distributed accelerator. Chromospheric HXR footpoints can show structure on arcsecond scales \citep[e.g.][]{2009ApJ...698.2131D} but tend not to be spatially resolved due to the restrictions of the imaging techniques involved. However, the strong association of HXR footpoints and optical footpoints \citep{2007ApJ...656.1187F} which can be extremely compact \citep{2007PASJ...59S.807I} implies that there is sub-resolution HXR structure present. The ability of RHESSI to produces images in many different energy intervals, and also the development of visibility-based imaging analysis, borrowed from radio astronomy, enables the vertical structure of the HXR emission region to be probed. The source centroid locations are found to vary systematically with energy with more energetic sources found deeper in the chromosphere \citep{2002SoPh..210..383A} as predicted by the collisional thick target, and a model of sub-resolution chromospheric strands each with a different density profile has been proposed to explain the thicker-than-expected vertical profile of HXR emission \citep{2010ApJ...717..250K}.  Evidence for electron beaming seems to be absent \citep[e.g.][]{2006ApJ...653L.149K}.

The association of optical and X-ray sources has been remarked upon earlier. Though quiet-sun optical radiation originates of course from the photosphere, it seems unfeasible to explain the optical enhancements as direct photospheric heating by electrons, due to the very large column depths that  these would have to traverse. Heating by protons - which travel much deeper than electrons with the same speed - remains a possibility, but so little is known about the spatial and temporal evolution of flare $\gamma$-ray sources that this is difficult to rule in or out.

However, it seems clear that at least in the flare impulsive phase the optical emission is closely related to the high numbers of non-thermal electrons present. A reasonable, but not the sole, hypothesis is that the optical emission is free-bound hydrogen recombination continuum generated following the ionisation of partially neutral portions of the chromosphere by flare-accelerated electrons \citep{1972SoPh...24..414H}. The UV-EUV component of this free-bound emission radiated towards the photosphere may also lead to enhanced photospheric heating and increased opacity (via enhanced ionisation of metals and, thereby, production of $H^-$) contributing also a photospheric component \citep{1989SoPh..124..303M}.  This is known as `backwarming'. Further progress on understanding the process requires observations of the flare optical continuum, particularly the hydrogen Balmer and Paschen continua. The hydrogen Paschen continuum has rarely been observed in solar flares \citep{1984SoPh...92..217N}, and broad-band observational study of the Balmer continuum \citep{1983SoPh...85..285N} has also been neglected in flares since the 1980s. A serendipitous observations in the Lyman continuum was made by SOHO/SUMER \citep{2004A&A...418..737L}, however recently the Solar Dynamics Observatory's Extreme Ultraviolet Variability Experiment (EVE) has been used to study the evolution of the Hydrogen Lyman continuum as well as the Helium I and II  Lyman continua during flares \citep{milliganetal2012}. With the availability of such consistent data, sampled at 10s cadence, we expect that substantial progress can be made on understanding the hydrogen free-bound continuum.

\begin{figure}
   \centering
  \includegraphics[width=0.9\textwidth]{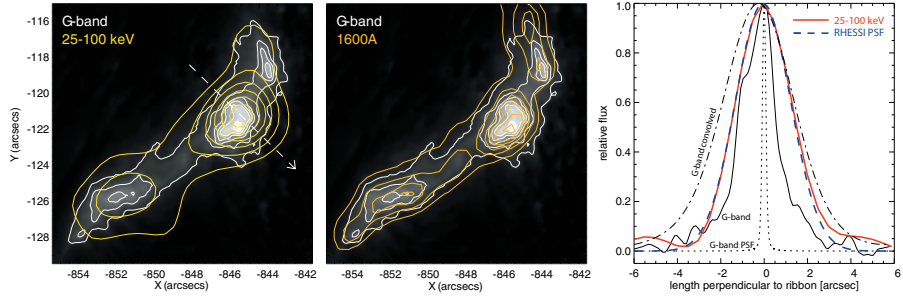}
  \caption{The \hinode X6.5 flare SOL2006-12-06T18:47 studied by \cite{2011ApJ...739...96K} Left: G-band difference image (white) with RHESSI 25-100~keV overlaid. Center: \hinode G-band difference image with TRACE 1600\AA\ contour overlaid.  Right: profiles taken along the direction of the arrow, demonstrating that the RHESSI footpoint profile is as wide as the RHESSI point-spread-function (thus is unresolved) but is wider than the equivalent G-band image when convolved with the RHESSI PSF. } 
             \label{fig:krucker}%
   \end{figure}

The \hinode solar optical telescope (SOT) has made a number of flare observations, which have added to our overall appreciation and understanding of white light flare processes. The extreme narrowness in places of the ribbons viewed in that part of the spectrum normally dominated by the CH bandhead (G-band) has been pointed out by \cite{2007PASJ...59S.807I}. The spatial relationships and overall energetics of the G-band, in comparison to HXRs has been studied by \cite{2010ApJ...715..651W} and \cite{2011ApJ...739...96K}.  A limited number of observations of flare footpoints in the SOT red, green and blue broad filters also exist (see K. Watanabe et al., this volume) and can be exploited to learn about the physical properties of optical footpoints. It should be noted that it is the combination of SOT G-band observations and RHESSI imaging spectroscopy that have put the firmest constraints so far of the properties of the non-thermal electron distribution in the flare chromosphere \citep{2011ApJ...739...96K}. As described in Section~\ref{sect:models}, the non-thermal chromospheric electron distribution required to produce the HXR emission in the  unresolved HXR footpoint of a major \hinode flare (Fig. \ref{fig:krucker}) is difficult to accommodate in the standard electron beam model, and this important result should encourage exploitation of SOT for targeted flare studies in the remainder of the current cycle.

Alluded to above, but not explained, is the question of why the UV, EUV and H$\alpha$ emission in flares is organised into ribbons, while HXRs and optical continuum tends to be concentrated into a few compact sources. The RHESSI dynamic range is of course limited, to a factor 5-10 or so, but nonetheless there are particular locations in the flare magnetic structure which receive a higher energy flux from the corona. The only theory advanced so far on this topic is that of \cite{2009ApJ...693.1628D} who associate the locations and motions of HXR footpoints with those of the photospheric endpoints of magnetic separators.  This is also an area that could be profitably studied using SOT spectropolarimetric and imaging data. More speculative,  \cite{2009ApJ...695.1151B}, who studied the enthalpy flux and the Poynting flux to the chromosphere during a flare using 3D MHD simulations, show that the regions of high enthalpy flux adopt a ribbon-like configuration, with a few spots of high Poynting flux having a footpoint-like appearance.

\subsection{EUV footpoints}\label{sect:EUV}
\begin{figure}
   \centering
  \includegraphics[width=0.9\textwidth]{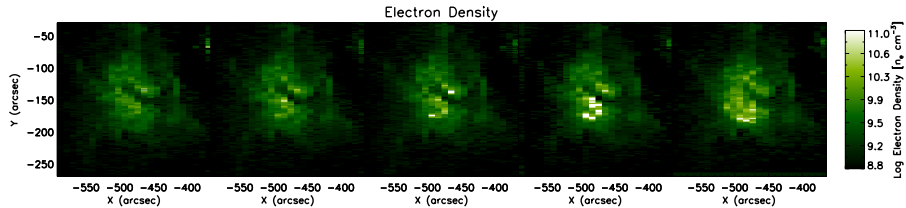}
  \caption{Density maps made with \hinode EIS, at several times during a small flare, using the Fe XIII $\lambda\lambda 203.8/\lambda\lambda 202.04$ diagnostic. Bright (i.e. high density) spots correspond to the flare ribbons, as seen by the TRACE satellite \citep{2011A&A...532A..27G}. } 
             \label{fig:graham}%
   \end{figure}

Flare impulsive-phase footpoint sources have been observed in the EUV range by the SOHO Coronal Diagnostic Spectrometer (CDS) in a small number of events \cite{2001ApJ...555..435B,2006ApJ...638L.117M,2006SoPh..234...95D}, which concentrated on footpoint line-of-sight velocity measurements, and also by the \hinode EUV Imaging Spectrometer (EIS). With EIS, density diagnostics of the hot footpoint plasma have been possible for the first time \citep{2010ApJ...719..213W,2011A&A...526A...1D,2011A&A...532A..27G,2011ApJ...740...70M}, and electron number densities up to and exceeding $10^{11}{\rm cm}^{-3}$ have been found using diagnostic line pairs formed at 1-1.5~MK (Fig. \ref{fig:graham}. Accompanying these high densities are high non-thermal speeds, up to 80~km/s, deduced from the broader-than-expected spectral line profiles, and both plasma upflows (evaporation) and downflows from the line centroids. The presence of hot, dense, possibly turbulent plasma at flare footpoints is not particularly surprising, having already been strongly suggested by analysis of imaging observations \citep{2004A&A...415..377M,2011ApJ...734...34K} but the EIS spectroscopic density diagnostics confirm this, and allow us to obtain a good observational understanding of the condition of the flare footpoint plasma.  The origin of the non-thermal speeds is still unclear, though \cite{2011ApJ...740...70M} points out that the interpretation of the line-broadenings as being `non-thermal' depends on the plasma being in LTE and ionisation equilibrium. That is, if a spectral line ordinarily formed in a plasma at 2MK is instead formed in a plasma with an electron temperature of 8~MK but in which the ionisation state has not yet `caught up,' with the electron temperature then the  broad line profiles might be readily explained.

 High temperature evaporative upflows, due to rapid heating of the chromosphere leading to explosive expansion, are expected. The downflows, when observed in the past as asymmetries in the H$\alpha$ line \citep[e.g.][]{1983SoPh...83...15T}, are interpreted as the momentum-conserving counterpart \citep{1990ApJ...348..333C}. What is a surprising in the current observations is that chromospheric impulsive phase downflows are observed to be present in lines with such high formation temperatures, up to 1.3~MK, or higher (T. Watanabe, in this volume). Radiation hydrodynamics simulations of electron beam energy deposition in the chromosphere show that the beam energy is deposited primarily mid-chromosphere resulting in a hot upflowing region, and a much cooler  ($\sim 10^4$ K) downwards propagating `condensation' wave. In their beam-driven simulations \cite{2009ApJ...702.1553L} can obtain downflowing heated chromospheric plasma, at temperatures of around 1MK, but these latter simulations only account for optically thin radiative losses, ignoring optically-thick emission and the time evolution of target ionization, which are both core to the hydrodynamic evolution \citep{2005ApJ...630..573A}. Thus \cite{2009ApJ...702.1553L} only cover cases where the bulk of the energy is deposited in an already optically-thin, fully-ionized chromosphere - or the upper chromosphere. It should also be pointed out that none of the published simulations so far attain the energy input rates of $10^{12}{\rm erg~cm^{-2}s^{-1}}$ implied by the observations of \cite{2011ApJ...739...96K}, but offer many interesting insights nonetheless.

%One thing to note is that in the quiet Sun, the chromosphere at a density of upwards of $10^{11}{\rm cm^{-3}}$ is not fully ionised. Since plasma at this density shows temperatures upwards of 1MK, the implication is that the ionisation state of the chromosphere is changed. This is not surprising. Also, at temperatures of 1MK, the plasma will not be in the over-ionised state, necessary to produce an enhanced optical continuum, but will pass through this regime as the flare heating dies away.

It is clear that more can be done with the chromospheric diagnostics available from EIS. In particular, work is underway on understanding the differential emission measure of flare footpoints, giving a first glimpse at the chromospheric plasma temperature distribution during a flare.  Not only EIS, but also SDO/EVE, will be valuable in this, though only spatially-integrated properties will be obtained with the Sun-as-a-star observations provided by the latter. One final remark is that interpretation of all of the EUV spectroscopic diagnostics described above rely to some extent at least on the `coronal approximation' assumptions: optically thin Maxwellian plasma in LTE and ionisation equilibrium. The energy flux radiated during a solar flare is at least an order of magnitude, perhaps two, greater than the quiescent energy flux of the Sun, so a flare is a very large and rapid perturbation to the equilibrium state. There is no guarantee that the conditions under which we can straightforwardly apply these familiar diagnostics will obtain in the flare chromosphere, so it is clear that more modeling of the validity of diagnostics under different conditions must be carried out \citep[e.g.][]{2004A&A...425..287B,2010SoPh..263...25D,2011A&A...531A.111D}.

\section{Future prospects}
As is described briefly above, there has been substantial progress on many fronts during the last two decades, but many gaps in our knowledge remain. In particular, the understanding of the flare chromosphere in the UV and optical has fallen behind compared to other wavelength ranges, and the anticipated future focus on chromospheric observations in the UV by the IRIS satellite and by the proposed SOLAR-C configuration, and in the optical by the Advanced Technology Solar Telescope towards the end of the decade, is to be welcomed also for flare studies. The flare chromosphere is the source of the majority of flare radiation, and the location where most of the flare energy ultimately dissipates, so it is with good reason that we should look here. When compared with our knowledge of stellar flares our knowledge of the chromospheric flare spectrum in the UV and optical is very poor (with the exception of H$\alpha$). The dearth of knowledge is understandable - it is difficult to catch a flare in the act with a slit spectrometer, especially when the size of the field of view has been reduced to enable increased spatial or temporal resolution.  But on the other hand the progress made in the EUV shows that flare imaging spectroscopy observations are possible and worthwhile, and we can anticipate progress in UV diagnostics of the heated chromosphere. Smaller instruments such as IBIS and ROSA working in the optical range, and with the very high cadence necessary to capture flare dynamics, pave the way for studies with the ATST. The Solar Orbiter mission, though restricted by telemetry and by available modes of operation during the long intervals between commanding, also offers excellent prospects for continued studies of flares in the EUV (SPICE, EUI - including also a Lyman $\alpha$ channel), optical (PHI) and HXR (STIX), with of course the added advantage of simultaneous in situ observations of flare-generated accelerated particles and plasma waves. And finally, observations from ALMA can be expected to show the far infrared and submillimeter emission from hot chromospheric flare plasma and from non-thermal electrons \citep{2009CEAB...33..309K}.

The condition of the flare plasma represents a severe challenge to modeling. The energy input to the flare chromosphere can be in excess of ten times the photospheric luminosity, representing a very severe perturbation to its equilibrium. There is a substantial non-thermal `tail' population of non-thermal electrons, as shown by the HXR observations, meaning that the effect on familiar diagnostic lines of non-thermal excitation, and also non-thermal ionisation must be calculated. An equilibrium, either coronal or Saha is unlikely to exist in the chromosphere at such times. In the hotter, less dense upper chromosphere the electron  and ion populations are unlikely to be collisionally coupled on the rapid flare timescales, meaning that at least a 2-fluid description of the chromospheric plasma is required. 

\acknowledgements The author is most grateful to the organisers of the Hinode-5 meeting for financial support and would also like to thank H. Hudson for useful comments.This work is supported by STFC grant ST/1001808/1 from the UK's Science and Technology Facilities Council, the EC-funded FP7 HESPE network and by Leverhulme Foundation Grant F00-179A. 

\bibliography{fletcher}

\end{document}